\documentstyle[aps,prd,epsfig,preprint]{revtex}
\def\fun#1#2{\lower3.6pt\vbox{\baselineskip0pt
\lineskip.9pt
\ialign{$\mathsurround=0pt#1\hfill##\hfil$\crcr#2
\crcr\sim\crcr}}}
\begin{document}
\vspace{0.5in}
\title{\vskip-2.5truecm{\hfill \baselineskip 14pt 
{\hfill {{\small \hfill UT-STPD-3/99 }}}\vskip .1truecm} 
\vspace{1.0cm}
\vskip 0.1truecm {\bf Leptogenesis in Supersymmetric 
Hybrid Inflation}}
\vspace{1cm}
\author{{G. Lazarides}\thanks{lazaride@eng.auth.gr}} 
\vspace{1.0cm}
\address{{\it Physics Division, School of Technology, 
Aristotle University of Thessaloniki,
\\ Thessaloniki GR 540 06, Greece}}
\maketitle

\vspace{2cm}

\begin{abstract}
\baselineskip 12pt

\par
The nonsupersymmetric as well as the supersymmetric hybrid 
inflationary model is reviewed. The scenario of baryogenesis 
via a primordial leptogenesis is discussed and the role of 
the nonperturbative electroweak sphaleron effects is 
analyzed in detail. A supersymmetric model based on a 
left-right symmetric gauge group, which `naturally' leads 
to hybrid inflation, is presented. The $\mu$ problem is 
solved, in this model, and the baryon asymmetry of 
the universe is produced through leptogenesis. For 
masses of $\nu_{\mu}$, $\nu_{\tau}$ from the small angle 
MSW resolution of the solar neutrino problem and 
SuperKamiokande, maximal $\nu_{\mu}-\nu_{\tau}$ mixing 
can be achieved. The required values of the relevant 
parameters are, however, quite small.
\end{abstract}

\thispagestyle{empty}
\newpage
\pagestyle{plain}
\setcounter{page}{1}
\def\beq{\begin{equation}}
\def\eeq{\end{equation}}
\def\beqa{\begin{eqnarray}}
\def\eeqa{\end{eqnarray}}
\def\tr{{\rm tr}}
\def\x{{\bf x}}
\def\p{{\bf p}}
\def\k{{\bf k}}
\def\z{{\bf z}}
\baselineskip 20pt

\section{Introduction} 
\label{sec:intro}

\par
The hybrid inflationary scenario \cite{hybrid}, which can 
reproduce the measurements of the cosmic background explorer 
(COBE) \cite{cobe} with more or less `natural' 
values of the relevant coupling constants, is almost 
automatically realized \cite{lyth,dss} in supersymmetric 
grand unified theories (GUTs). In particular, 
a moderate extension of the minimal supersymmetric standard 
model (MSSM) based on a left-right symmetric gauge group 
provides \cite{lss} a `natural' framework for the 
implementation of hybrid inflation. The $\mu$ problem of 
MSSM can be easily resolved \cite{dls} in the context 
of this model by coupling the inflaton system to the 
electroweak higgs superfields.

\par
At the end of inflation, the inflaton (oscillating system) 
predominantly decays into electroweak higgs superfields, 
thereby `reheating' the universe. However, its subdominant 
decay mode to right handed neutrinos leads \cite{atmo}, 
via their subsequent decay, to the production of a 
primordial lepton asymmetry in the universe. Nonperturbative 
electroweak sphaleron effects, which violate baryon and 
lepton number, then partially transform this asymmetry to 
the observed baryon asymmetry of the universe (BAU).

\par
We analyze the consequences of this baryogenesis 
mechanism on $\nu_{\mu}-\nu_{\tau}$ mixing. 
We find \cite{atmo} that, for masses of $\nu_{\mu}$, 
$\nu_{\tau}$ which are consistent with the small angle 
MSW resolution of the solar neutrino problem and the 
recent results of the SuperKamiokande experiment 
\cite{superk}, maximal $\nu_{\mu}-\nu_{\tau}$ 
mixing can be achieved. The required values of the 
relevant parameters are, however, quite small.

\par
In Sec.\ref{sec:hybrid}, we review the nonsupersymmetric 
(Sec.\ref{subsec:nonsusy}) as well as the supersymmetric 
(Sec.\ref{subsec:susy}) version of the hybrid 
inflationary scenario. In Sec.\ref{sec:baryons}, we discuss
baryogenesis through a primordial leptogenesis. In 
particular, Sec.\ref{subsec:leptons} is devoted to the 
generation of the primordial lepton number. The 
topologically nontrivial structure of the vacuum in gauge 
theories and the resulting nonperturbative baryon and 
lepton number violating phenomena in the standard model 
are analyzed in Sec.\ref{subsec:sphaleron}. The rate of 
these phenomena at finite temperatures is calculated by 
employing electroweak sphalerons and the final BAU is 
estimated. Finally, in Sec.\ref{sec:lr} the supersymmetric 
model based on a left-right symmetric gauge group is 
presented. In particular, the solution of the $\mu$ problem 
(Sec.\ref{subsec:mu}), inflation (Sec.\ref{subsec:inf}) 
and leptogenesis (Sec.\ref{subsec:reheatlepto}) are sketched.

\section{Hybrid Inflation} 
\label{sec:hybrid}

\subsection{The non Supersymmetric Version} 
\label{subsec:nonsusy}

\par
The most important disadvantage of most inflationary 
scenarios was that they needed extremely small coupling 
constants in order to reproduce the results of 
COBE \cite{cobe}. This difficulty 
was overcome some years ago by Linde \cite{hybrid} who 
proposed, in the context of nonsupersymmetric GUTs, 
a clever inflationary  scenario known as 
hybrid inflation. The idea was to use two real scalar fields 
$\chi$ and $\sigma$ instead of one that was normally used. 
The field $\chi$ provides the vacuum energy which drives 
inflation while  $\sigma$ is the slowly  varying  field 
during inflation. The main advantage of this scenario is 
that it can reproduce the observed temperature fluctuations 
of the cosmic background radiation (CBR) with `natural' 
values of the parameters in contrast to previous 
realizations of inflation (like the `new' \cite{new} or 
`chaotic' \cite{chaotic} inflationary scenarios). The 
potential utilized by Linde is
\begin{equation}
V ( \chi, \sigma)= \kappa^2 \left( M^2 - 
\frac {\chi^2}{4}\right)^2 +   \frac{\lambda^2 
\chi^2 \sigma^2}   {4}   +   \frac {m^2\sigma^2}{2}~~,
\label{eq:lindepot}
\end{equation}
where $\kappa,~\lambda$ are dimensionless positive 
coupling constants and $M$, $m$ mass parameters. The 
vacua lie  at $\langle \chi\rangle=\pm 2 M$, $\langle 
\sigma \rangle=0$. Putting $m$=0, for the moment, we 
observe that the potential possesses an exactly flat 
direction at $\chi=0$ with $V(\chi=0 ,\sigma)=\kappa^2 
M^4$. The mass squared of the field $\chi$ along this flat 
direction is given by $m^2_\chi=-\kappa^2 M^2+\frac{1}{2}
\lambda^2 \sigma^2$ and remains nonnegative for $\sigma 
\geq \sigma_c = \sqrt{2}\kappa M/\lambda $. This means 
that, at $\chi=0$ and $\sigma \geq \sigma_c$, we obtain 
a valley of minima with  flat bottom. Reintroducing the 
mass parameter $m$ in Eq.(\ref{eq:lindepot}), we observe 
that this  valley acquires a nonzero slope. A region of the 
universe, where $\chi$ and $\sigma$ happen to be almost 
uniform with negligible kinetic energies and with values 
close to the bottom of the valley of minima, follows this 
valley in its subsequent evolution and undergoes inflation. 
The quadrupole anisotropy of CBR produced during this 
inflation can be estimated to be
\begin{equation}
\left(\frac {\delta T}{T}\right)_{Q} 
\approx  \left(\frac {16 \pi}{45}\right)^{1/2} 
\frac{\lambda \kappa^2 M^5}
{M^3_Pm^2}~,
\label{eq:lindetemp}
\end{equation}
where $M_{P}=1.22\times 10^{19}{\rm{GeV}}$ is the Planck
scale. The COBE~\cite{cobe} result, $(\delta T/T)_{Q} 
\approx 6.6 \times 10^{-6}$, can then be reproduced with 
$M\approx 2.86\times 10^{16}$ GeV, the supersymmetric GUT 
vacuum expectation value (vev), and $m \approx 1.3~\kappa 
\sqrt {\lambda}\times 10^{15}$ GeV $\sim 10^{12}$ GeV 
for $\kappa, \lambda \sim 10^{-2}$. Inflation terminates 
abruptly at $\sigma=\sigma_{c}$ and is followed by a 
`waterfall', i.e., a sudden entrance into an oscillatory 
phase about a global minimum. Since the system can fall into 
either of the two available global minima with equal 
probability, topological defects are copiously produced if 
they are predicted by the particular particle physics model 
one is considering.

\subsection{The Supersymmetric Version}
\label{subsec:susy}

\par
The hybrid inflationary scenario is \cite{lyth} `tailor 
made' for application to supersymmetric GUTs except that 
the mass of $\sigma$, $m$, is unacceptably large for 
supersymmetry, where all scalar fields acquire masses of 
order $m_{3/2} \sim 1$ TeV (the gravitino mass) from 
soft supersymmetry breaking. To see this, consider a 
supersymmetric GUT with a (semi-simple) gauge group $G$ of 
rank $\geq 5$ with $G \to G_S$ (the standard model gauge 
group) at a scale $M \sim 10^{16}$ GeV. The spectrum of 
the theory below $M$ is assumed to coincide with the 
MSSM spectrum plus standard 
model singlets so that the successful predictions for 
$\alpha_{s}$, ${\rm{sin}}^{2} \theta_{W}$ are retained. 
The theory may also possess global symmetries. The breaking 
of $G$ is achieved through the superpotential
\begin {equation}
W =\kappa S( \phi\bar{\phi}- M^2),
\label{eq:superpot}
\end {equation}
where $\phi$, $\bar{\phi}$ is a conjugate pair of 
standard model singlet left handed superfields which belong 
to nontrivial representations of $G$ and reduce its rank 
by their vevs and $S$ is a gauge singlet left handed 
superfield. The coupling constant $\kappa$ and the mass 
parameter $M$ can be made positive by phase redefinitions. 
This superpotential is the most general renormalizable 
superpotential consistent with a $U(1)$ R-symmetry under 
which $W \to e^{i\theta} W,~S \to e^{i \theta}S,~\phi 
\bar {\phi} \to \phi\bar{\phi}$ and gives the potential
\begin{equation}
V=\kappa^2 \mid M^2-\phi\bar{\phi}\mid^2 
 +\kappa^2 \mid S \mid^2
(\mid \phi \mid^2 +\mid \bar{\phi}\mid^2)
+{\rm{ D-terms}}.
\label{eq:hybpot}
\end{equation}
Restricting ourselves to the D flat direction $\phi=
\bar{\phi}^* $ which contains the supersymmetric minima and 
performing appropriate gauge and R- transformations, we can 
bring $S$, $\phi$, $\bar{\phi}$ on the real axis, i.e., 
$S \equiv \sigma/\sqrt{2}$, $\phi=\bar{\phi} \equiv 
\chi/2$, where $\sigma$, $\chi$ are normalized real scalar 
fields. The potential then takes the form in 
Eq.(\ref{eq:lindepot}) with $\kappa = \lambda$ and $m=0$ and, 
thus, Linde's potential for hybrid inflation is almost
obtainable from supersymmetric GUTs but without the mass 
term of $\sigma$ which is, however, of crucial importance 
since it provides the slope of the valley of minima 
necessary for inflation.

\par
One way to obtain a valley of minima useful for inflation 
is \cite{lp} to replace the trilinear term 
in $W$ in Eq.(\ref{eq:superpot}) by 
the next order nonrenormalizable coupling. Another way, 
which we will adopt here, is \cite{dss} to keep the 
renormalizable superpotential in Eq.(\ref{eq:superpot}) 
and use the radiative corrections along the inflationary 
valley ($\phi=\bar{\phi}= 0$~, $S > S_{c}\equiv M$).
In fact, due to the mass splitting in the supermultiplets 
$\phi$, $\bar{\phi}$ caused by the supersymmetry 
breaking `vacuum' energy density $\kappa^{2} M^{4}$ 
along this valley, there are important radiative corrections. 
At one-loop, the inflationary potential is given 
\cite{dss,lss} by
$$ 
V_{\rm{eff}}(S) = \kappa^{2}M^{4}
$$
\begin{equation} 
\left[
1 +\frac{\kappa^2}{32 \pi^2} \left(2\ln\left(
\frac{\kappa^{2}S^{2}}{\Lambda^2}\right)+ 
\left(\frac{S^2}{S_{c}^2}-1\right)^2 
\ln\left(1-\frac{S_{c}^2}{S^2}\right) 
+\left(\frac{S^2}{S_{c}^2}+1\right)^2 
\ln\left(1+\frac{S_{c}^2}{S^2}\right)\right)  
\right]~,
\label{eq:veffexact}
\end{equation}
where $\Lambda$ is a suitable mass renormalization scale.
For $S$ sufficiently larger than $S_{c}$,
\begin{equation}
V_{{\rm{eff}}} (S) = \kappa^2 M^4 
\left[ 1 + \frac{\kappa^2}{16\pi^2} \left( \ln
\left(\frac{\kappa^{2} S^{2}}{\Lambda^2}\right) + 
\frac{3}{2} - \frac{S_c^4}{12S^4} + 
\cdots \right)\right]~.
\label{eq:veff}
\end{equation}
Using this effective potential, one finds 
that the cosmic microwave quadrupole anisotropy
\begin{equation} 
\left(\frac{\delta T}{T}\right)_{Q} \approx 8 \pi 
\left(\frac{N_{Q}}{45}\right)^{1/2} 
\frac{x_{Q}}{y_{Q}}\left(\frac{M}{M_{P}}\right)^2~.
\label{eq:cobe}
\end{equation}
Here, $N_{Q}$ is the number of e-foldings experienced by 
the universe between the time the quadrupole scale exited 
the inflationary horizon and the end of inflation and 
$y_{Q}=x_{Q}(1-7/(12x_{Q}^2)+\cdots)$ with 
$x_{Q}=S_{Q}/M$, $S_{Q}$ being the value of the scalar 
field $S$ when the scale which evolved to the present 
horizon size crossed outside the de Sitter (inflationary) 
horizon. Also, from 
Eq.(\ref{eq:veff}), one finds
\begin{equation} 
\kappa \approx \frac{8\pi^{3/2}}{\sqrt{N_{Q}}}
~ y_{Q}~\frac{M}{M_{P}}~.
\label{eq:kappa}
\end{equation}  

\par
The inflationary phase ends as $S$ approaches $S_{c}$ from 
above. Writing $S=xS_{c}$, $x=1$ corresponds to the phase 
transition from $G$ to $G_{S}$ which, as it turns out, 
more or less coincides with the end of the inflationary 
phase as one deduces from the slow roll conditions 
\cite{dss,lazarides}. Indeed, the $50-60$ e-foldings needed 
for the inflationary scenario can be realized even with small 
values of $x_{Q}$. For definiteness, we will take 
$x_{Q}\approx 2$. From COBE \cite{cobe} one then obtains 
$M \approx 5.5 \times 10^{15}~{\rm{GeV}}$ and 
$\kappa\approx 4.5\times 10^{-3}$ for $N_{Q}\approx 56$. 
Moreover, the primordial density fluctuation spectral index 
$n \simeq 0.98$. We see that the relevant part of inflation 
takes place at $S\sim 10^{16}~{\rm{GeV}}$. An interesting 
consequence of this is \cite{lyth,lss,sugra} that the 
supergravity corrections can be negligible.

\par
In conclusion, it is important to note that the superpotential 
$W$ in Eq.(\ref{eq:superpot}) leads to hybrid inflation in a 
`natural' way. This means that a) there is no need of very 
small coupling constants, b) $W$ is the most general 
renormalizable superpotential allowed by the gauge and R- 
symmetries, c) supersymmetry guarantees that the radiative 
corrections do not invalidate inflation, but rather provide 
a slope along the inflationary trajectory which drives the 
inflaton towards the supersymmetric vacua, and 
d) supergravity corrections can be brought under control
 so as to leave inflation intact.

\section{Baryogenesis via Leptogenesis} 
\label{sec:baryons}

\subsection{Primordial Leptogenesis} 
\label{subsec:leptons}

\par
In most hybrid inflationary models, it is not convenient to 
produce the observed BAU in the customary way, i.e., through 
the decay of color $~3,~\bar{3}~$ fields $(g,~g^{c})$. 
Some of the reasons are the following: i) For theories where 
leptons and quarks belong to different representations of 
the unifying gauge group $G$ (which is the case, for example, 
for $G=G_{LR}\equiv SU(3)_c\times SU(2)_L\times SU(2)_R
\times U(1)_{B-L}$ or $SU(3)_c \times SU(3)_L \times 
SU(3)_R$), the baryon number can be made almost 
exactly conserved by imposing an appropriate discrete 
symmetry. In particular, for $G=G_{LR}$, we can impose 
\cite{baryonparity} a discrete symmetry under which 
$q\rightarrow - q$, $q^{c} \rightarrow - q^{c}$, 
$\bar{q} \rightarrow - \bar{q}$, $\bar{q}^{c} 
\rightarrow - \bar{q}^{c}$ and all other superfields
remain invariant ($q,~q^{c},~\bar{q},~\bar{q}^{c}$ 
are superfields with the quantum numbers of the quarks, 
antiquarks and their conjugates respectively). ii) For 
theories where such a discrete symmetry is absent, we 
could, in principle, use as inflaton a pair of conjugate 
standard model singlet superfields $N$, $\bar{N}$ which 
decay into $g$, $g^{c}$. For $G=SU(3)_c \times SU(3)_L 
\times SU(3)_R$, $N$ ($\bar{N}$) could be the standard 
model singlet component of the (1, $\bar{3}$, 3) 
( (1, 3, $\bar{3}$) ) superfields with zero $U(1)_{B-L}$ 
charge. But this is again unacceptable since the breaking 
of $SU(3)_c \times SU(3)_L \times SU(3)_R$ by the vevs 
of $N$, $\bar{N}$ predicts \cite{trinifiedmonopoles} 
magnetic monopoles which can then be copiously produced 
after inflation. Also the gravitino constraint 
\cite{gravitino} on the `reheat' temperature, $T_{r} 
\stackrel{_{<}}{_{\sim }} 10^9$ GeV, implies $m_g 
\stackrel{_{<}}{_{\sim }} 10^{10}$ GeV (~from the 
coupling ($~m_g / \langle N\rangle) Ngg^{c}$~) 
leading to strong deviation from MSSM and possibly 
proton decay.

\par
So it is preferable to produce first a primordial lepton 
asymmetry \cite{lepto} which can then be partially turned 
into the observed baryon asymmetry of the universe by the 
nonperturbative sphaleron effects \cite{sphaleron} of 
the electroweak sector. In the particular model based on 
$G_{LR}$ which we will consider later, this is the only 
way to produce the BAU since the inflaton decays into 
higgs superfields and right handed neutrinos. The subsequent 
decay of right handed neutrinos into ordinary higgs 
particles (higgsinos) and light leptons (sleptons) can 
produce the primordial lepton asymmetry. It is important, 
though, to ensure that this primordial lepton asymmetry 
is not erased \cite{turner} by lepton number violating 
$2 \rightarrow 2$ scattering processes such as 
$ l l \rightarrow h^{(1)}\,^{*} h^{(1)}\,^{*}$ or 
$l h^{(1)} \rightarrow \bar{l} h^{(1)}\,^{*}$ ($l$ 
represents a lepton doublet and $h^{(1)}$ the higgs 
superfield which couples to the up type quarks) at all 
temperatures between $T_{r}$ and 100 GeV. This is 
automatically satisfied since the primordial lepton 
asymmetry is protected \cite{ibanez} by supersymmetry 
at temperatures between $T_{r}$ and $T \sim 10^{7}$ GeV, 
and for $T \stackrel{_{<}}{_{\sim }} 10^{7}$ GeV, 
these $2 \rightarrow 2$ scattering processes are well 
out of equilibrium provided \cite{ibanez} 
$m_{\nu_{\tau}}\stackrel{_<}{_\sim} 10~{\rm{eV}}$, 
which readily holds in our case (see below). 

\par
The lepton asymmetry produced by the out-of-equilibrium 
decay ($M_{\nu^{c}_{i}} \gg T_{r})$  of the right 
handed neutrinos $\nu^{c}_{i}$, which emerged from the 
inflaton decay, is \cite{lepto}
\begin{equation}
\frac {n_{L}}{s} \approx -  \frac{3}{16 \pi} 
\frac {T_{r}}{m_{infl}}\sum_{l \neq i}
g(r_{li}) \frac{{\rm{Im}}(U~M^{D}\,^{\prime}
~M^{D}\,^{\prime}\,^{\dagger}~U ^{\dagger})^{2}_{il}}
{|\langle h^{(1)}\rangle|^{2}(U~M^{D}\,^{\prime}
~M^{D}\,^{\prime}\,^{\dagger}~U ^{\dagger})_{ii}}~,
\label{eq:genlept}
\end{equation}
where $n_L$ and $s$ are the lepton number and entropy 
densities, $m_{infl}$ in the inflaton mass, 
$M^{D}\,^{\prime}$ is the diagonal `Dirac' 
mass matrix, $U$ a unitary transformation so that 
$UM^{D}\,^{\prime}$ is the `Dirac' mass matrix in the 
basis where the `Majorana' mass matrix of $\nu^{c}$~'s 
is diagonal and $|\langle h^{(1)} \rangle| \approx 
174~{\rm{GeV}}$ for large ${\rm tan}\beta$. The function
\begin{equation}
g(r_{li}) = r_{li}~\ln \left(\frac {1 + r^{2}_{li}}
{r^{2}_{li}}\right)~,~r_{li} = \frac {M_{l}} {M_{i}}~,
\label{eq:gfunction}
\end{equation}
with $g(r) \sim 1/r$ as $r \rightarrow \infty$.
Here we have taken into account the following prefactors: 
i) At `reheat', $n_{infl} m_{infl} =
(\pi^{2}/30) g_{*} T^{4}_{r}$ ($n_{infl}$ is the 
inflaton number density and $g_{*}$ the effective 
number of massless degrees of freedom) which together 
with the relation $s = (2 \pi^{2}/45) g_{*}T^{3}_{r}$
implies that $n_{infl}/s = (3/4)(T_{r}/m_{infl})$.
ii) Since each inflaton decays into two $\nu^{c}$~'s, 
their number density $n_{\nu^{c}} = 2 n_{infl}$ which 
then gives $n_{\nu^{c}}/s= (3/2)(T_{r}/m_{infl})$. 
iii) Supersymmetry gives an extra factor of two.

\subsection {Sphaleron Effects} 
\label{subsec:sphaleron}

\par
To see how the primordial lepton asymmetry partially turns 
into the observed BAU, we must first discuss the 
nonperturbative baryon ($B$-) and lepton ($L$-) number 
violation \cite{thooft} in the standard model. Consider the 
electroweak gauge symmetry $SU(2)_{L} \times U(1)_{Y}$ 
in the limit where the Weinberg angle $\theta_W=0$ and 
concentrate on $SU(2)_{L}$ (inclusion of $\theta_{W}\neq 0$ 
does not alter the conclusions). Also, for the moment, ignore 
the fermions and higgs fields so as to have a pure 
$SU(2)_{L}$ gauge theory. This theory has \cite{vacuum} 
infinitely many classical vacua which are topologically 
distinct and are characterized by a `winding number' 
$n \in Z$. In the `temporal gauge' ($A_{0}=0$), the 
remaining gauge freedom consists of time independent 
transformations and the vacuum corresponds to a pure gauge
\begin{equation}
A_{i} = \frac{i}{g}~\partial _{i}g(\bar{x}) 
g ^{-1}(\bar{x})~,
\label{eq:gauge}
\end{equation}
where $g$ is the $SU(2)_{L}$ gauge coupling constant, 
$\bar{x}$ belongs to ordinary 3-space, $i$ =1,2,3, 
$g(\bar{x}) 
\in SU(2)_{L}$, and $g(\bar{x}) \rightarrow 1 $ as 
$ \mid\bar{x}\mid \rightarrow \infty$. Thus, the 3-space 
compactifies to a sphere $S^{3}$ and $g(\bar{x})$ defines 
a map: $S^{3} \rightarrow SU(2)_{L}$ (with the 
$SU(2)_{L}$ group being topologically equivalent to 
$S^{3}$). These maps are classified into homotopy classes 
constituting the third homotopy group of $S^{3},~\pi_{3}
(S^{3})$, and are characterized by a `winding number'
\begin{equation}
n = \int d^{3}x~\epsilon^{ijk}~{\rm{tr}} 
\left(\partial_{i}g(\bar{x}) 
g^{-1}(\bar{x})\partial_{j}g(\bar{x}) g^{-1}(\bar{x})
\partial_{k}g(\bar{x}) g^{-1}(\bar{x})\right).
\label{eq:wind}
\end{equation}
The corresponding vacua are denoted as $\mid n\rangle$, 
$n\in Z$.

\par
The tunneling amplitude from the vacuum $\mid n_{-}
\rangle$ at $t=-\infty$ to the vacuum $\mid n_{+}
\rangle~$ at $t=+\infty$ is given by the functional 
integral
\begin{equation}
\langle n_{+}\mid n_{-}\rangle = \int(dA)~e^{-S(A)}
\label{eq:path}
\end{equation}
over all gauge field configurations satisfying the 
appropriate boundary conditions at $t=\pm \infty$.
Performing a Wick rotation, 
~$x_0 \equiv t \rightarrow -i x_{4}$, 
we can go to Euclidean space-time. Any Euclidean field 
configuration with finite action is characterized by an 
integer topological number known as the Pontryagin number
\begin{equation}
q = \frac {g^{2}}{16\pi^{2}} 
\int d^{4}x~{\rm{tr}}\left(F^{\mu \nu} 
\tilde{F}_{ \mu \nu}\right)~,
\label{eq:pontryagin}
\end{equation}
with $\mu$,$\nu$=1,2,3,4 and $\tilde{F}_{\mu \nu}=
\frac {1}{2}\epsilon_{\mu \nu \lambda \rho}
F^{\lambda \rho}$ being the  dual field strength. But 
${\rm{tr}} (F^{\mu \nu} \tilde{F}_{\mu \nu}) = 
\partial ^{\mu} J_{\mu}$, where $J_{\mu}$ is the 
`Chern-Simons current' given by
\begin{equation}
J_{\mu}=\epsilon_{\mu \nu \alpha \beta}~{\rm{tr}}
\left(A^{\nu}
F^{\alpha \beta}-\frac{2}{3}gA^{\nu}A^{\alpha}
A^{\beta}\right).
\label{eq:csc}
\end{equation}
In the `temporal gauge' ($A_0=0$),
\begin{eqnarray*}
q=\frac{g^{2}}{16 \pi^{2}} \int d^{4}x
~\partial^{\mu}J_{\mu}=
\frac{g^{2}}{16\pi^{2}}
\mathop{\Delta}_{x_{4}=\pm \infty} 
\int d^{3}x~J_{0}   
\end{eqnarray*}
\begin{equation}
=\frac{1}{24\pi^{2}}\mathop{\Delta}_{x_{4}=
\pm \infty} 
\int d^{3}x~\epsilon^{ijk}~ 
{\rm{tr}}\left(\partial_{i} g g^{-1}\partial_{j}
gg^{-1}\partial_{k}gg^{-1}\right)=n_{+}-n_{-}~.
\label{eq:interpol}
\end{equation}
This means that the Euclidean field configurations which 
interpolate between the vacua $\mid n_{+}\rangle,
~\mid n_{-}\rangle$ at $x_4 =\pm \infty$ have 
Pontryagin number $q= n_{+}-n_{-}$ and the  path integral 
in Eq.(\ref{eq:path}) should be performed over all these 
field configurations.

\par
For a given $q$, there is a lower bound on $S(A)$,
\begin{equation}
S(A) \geq \frac{8 \pi^{2}}{g^{2}}\mid q \mid~,
\label{eq:lbound}
\end{equation}
which is saturated if and only if $F_{\mu \nu}=
\pm \tilde{F} _{\mu \nu}$, i.e, if the 
configuration is self-dual or self-antidual. For $q$=1, 
the self-dual classical solution is called instanton 
\cite{instanton} and is given by (in the `singular' gauge)
\begin{equation}
A_{a \mu}(x)=\frac{2 \rho^{2}}{g(x-z)^{2}}
~\frac{\eta_{a \mu \nu} (x-z)^{\nu}}
{(x-z)^{2} + \rho^{2}}~,
\label{eq:instanton}
\end{equation}
where $\eta_{a \mu \nu}$ ($a$=1,2,3; $\mu$,$\nu$=
1,2,3,4) are the t' Hooft symbols with $\eta_{aij}=
\epsilon_{aij}$ ($i$,$j$=1,2,3), $\eta_{a4i}=
-\delta_{ai}$, $\eta_{ai4}=\delta_{ai}$ and 
$\eta_{a44}=0$. The instanton depends on four 
Euclidean coordinates $z_{\mu}$ (its position) and 
its scale (or size) $\rho$. Two successive vacua 
$\mid n\rangle$,~$\mid n+1\rangle$ are separated by 
a potential barrier of height $\propto \rho^{-1}$.
The Euclidean action of the interpolating instanton is 
always equal to $8 \pi^{2}/g^{2}$, but the height of 
the barrier can be made arbitrarily small since the size 
$\rho$ of the instanton can be taken arbitrarily large.

\par
We now reintroduce the fermions into the theory and 
observe~\cite{thooft} that the $B$- and $L$- number 
currents carry anomalies, i.e.,
\begin{equation}
\partial_{\mu} J^{\mu}_{B} = 
\partial _{\mu} J^{\mu}_{L} = 
- n _{g} \frac {g^{2}}{16 \pi ^{2}}~{\rm{tr}} 
(F_{\mu \nu} \tilde{F}^{\mu \nu})~,
\label{eq:anomaly}
\end{equation}
where $n_{g}$ is the number of generations. It is then 
obvious that the tunneling from $\mid n_{-}\rangle$ to  
$\mid n_{+}\rangle$ is accompanied by a change of the
$B$- and $L$- numbers, $\Delta B=\Delta L=- n_{g}q=
- n_{g} (n_{+}-n_{-})$. Note that i) $\Delta (B-L)=0$, 
and ii) for $q$=1, $\Delta B=\Delta L=-3$ which means 
that we have the annihilation of one lepton per family 
and one quark per family and color (12-point function).

\par
We, finally, reintroduce the Weinberg-Salam higgs doublet 
$h$ with its vev given by
\begin{equation}
<h> = \frac {v}{\sqrt{2}}~\left(\matrix{0 \cr 1 \cr}
\right)~,~v \approx 246 ~{\rm{GeV}}~.
\label{eq:vev}
\end{equation}
It is then easy to see that the instanton ceases to exist 
as an exact solution. It is replaced by the so called 
`restricted instanton' \cite{restricted} which is an 
approximate solution for $\rho \ll v^{-1}$. For 
$\mid x-z\mid \ll \rho$, the gauge field configuration 
of the `restricted instanton' essentially coincides with 
that of the instanton and the higgs field is
\begin{equation}
h(x) \approx \frac {v}{\sqrt{2}}
~\left(\frac{(x-z)^{2}}
{(x-z)^{2} + \rho^{2}}\right)^{1/2} 
\left(\matrix {0 \cr 1 \cr } \right)~~.
\label{eq:restricted}
\end{equation}
For $\mid x-z \mid \gg \rho$, the gauge and higgs fields 
decay to a pure gauge and the vev in Eq.(\ref{eq:vev}) 
respectively. The action of the `restricted instanton' is 
$S_{ri}=(8 \pi^{2}/g^{2})+\pi^{2} v^{2} \rho^{2}+
\cdots$, which implies that the contribution of big size 
`restricted instantons' to the path integral in 
Eq.(\ref{eq:path}) is suppressed. This justifies 
{\it a posteriori} the fact that we restricted ourselves 
to approximate instanton solutions with $\rho \ll v^{-1}$.

\par
The height of the potential barrier between the vacua 
$\mid n\rangle,~\mid n+1\rangle$ cannot be now 
arbitrarily small. This can be understood by observing 
that the static energy of the `restricted instanton' at 
$x_{4}=z_{4}$ ($\lambda$ is the higgs self-coupling),
\begin{equation}
E_{b}(\rho) \approx \frac{3 \pi^{2}}
{g^{2}}~\frac{1}{\rho} + 
\frac {3}{8}\pi^{2} v^{2} \rho^{2} +  
\frac {\lambda}{4}\pi^{2} v^{4}\rho^{3}~,
\label{eq:static}
\end{equation}
is minimized for
\begin{equation}
\rho_{{\rm{min}}} = \frac {\sqrt{2}}{gv}
\left(\frac {\lambda}
{g^{2}}\right)^{-1/2}\left(\left(\frac {1}{64} + 
\frac {\lambda}{g^{2}}\right)^{1/2} - 
\frac {1}{8}\right)^{1/2} 
\sim M^{-1}_{W}~,
\label{eq:rhomin}
\end{equation}
and, thus, the minimal height of the potential barrier 
turns out to be $E_{{\rm{min}}} \sim M_{W} / \alpha_W$ 
($M_{W}$ is the weak mass scale and 
$\alpha_{W}=g^{2}/4 \pi$).
The static solution which corresponds to the top 
(saddle point) of this potential barrier is called 
sphaleron \cite{sphaleronsol} and is given by 
\begin{equation}
\bar{A} = v~\frac {f(\xi)}{\xi}~\hat{r} 
\times \bar{\tau}~,
~h = \frac {v}{\sqrt{2}}~t(\xi)~ \hat {r}\cdot  
\bar{\tau} \left (\matrix{0 \cr 1 \cr} \right),
\label{eq:sphaleron}
\end{equation}
where $\xi=2M_{W}r$, $\hat{r}$ is the  radial unit vector 
in ordinary 3-space 
and the 3-vector $ \bar{\tau}$ consists of the Pauli 
matrices. The functions $f(\xi),~t(\xi)$, which can be 
determined numerically, tend to zero as $\xi \rightarrow 0$ 
and to 1 as $\xi \rightarrow \infty$. The mass (static 
energy) of the sphaleron solution is estimated to be
\begin{equation}
E_{{\rm{sph}}} = \frac {2M_{W}}{\alpha_{W}}~k,
~1.5 \leq k \leq 2.7~,~ {\rm{for}} ~0 \leq 
\lambda \leq \infty~,
\label{eq:sphmass}
\end{equation}
and lies between 10 and 15 TeV.

\par
At zero temperature the tunneling from $\mid n\rangle$ to 
$\mid n+1\rangle$ is utterly suppressed \cite{thooft} by 
the factor exp$(-8 \pi^{2} /g ^{2})$. At high temperatures, 
however, thermal fluctuations over the potential barrier are 
frequent and this transition can occur \cite{sphaleron} with 
an appreciable rate. For $M_{W} \stackrel{_{<}}{_{\sim }} 
T \stackrel{_{<}}{_{\sim }} T_{c}$ ($T_c$ is the 
critical temperature of the electroweak transition), this 
rate can be calculated \cite{sphaleron} by expanding around 
the sphaleron (saddle point) solution and turns out to be
\begin{equation}
\Gamma \approx 10^{4}~ n_{g}~ \frac {v (T)^{9}}{T^{8}}
~{\rm{exp}} (-E_{{\rm{sph}}}(T)/T)~.
\label{eq:sphrate}
\end{equation}
Assuming that the electroweak phase transition is a second 
order one, $v(T)$ and $E_{{\rm{sph}}}(T) \propto 
(1 - T^{2}/T^{2}_{c})^{1/2}$. One can then show that 
$\Gamma \gg H$ ($H$ is the Hubble parameter) for 
temperatures $T$ between $\sim 200$ 
GeV and $\sim T_{c}$. Furthermore, for temperatures above 
$T_{c}$, where the sphaleron solution ceases to exist, it 
was argued~\cite{sphaleron} that we still have $\Gamma 
\gg H$. The overall conclusion is that nonperturbative 
$B$- and $L$- number violating processes are in 
equilibrium in the universe for cosmic temperatures 
$T\stackrel{_{>}}{_{\sim }} 200$ GeV. Remember that 
$B-L$ is conserved by these processes.

\par
Given a primordial $L$- number density, one can 
calculate \cite{turner,ibanez} the resulting $n_{B}/s$ 
($n_B$ is the $B$- number density). In MSSM, the 
$SU(2)_{L}$ instantons produce the effective operator 
(in symbolic form)
\begin{equation}
O_{2} = (q q q l)^{n_{g}} ( \tilde {h}^{(1)} 
\tilde {h}^{(2)}) 
\tilde{W}^{4}~,
\label{eq:woperator}
\end{equation}
and the $SU(3)_{c}$ instantons the operator
\begin{equation}
O_3 = ( q q u^{c} d^{c})^{n_{g}} \tilde {g}^{6}~,
\label{eq:coperator}
\end{equation}
where $q$, $l$ are the quark, lepton $SU(2)_{L}$ doublets 
respectively, $u^{c}$, $d^{c}$ the up, down type antiquark 
$SU(2)_{L}$ singlets respectively, $h^{(1)}$, $h^{(2)}$
the higgses which couple to up, down type quarks respectively, 
$g$, $W$ the gluons and $W$ bosons and tilde represents 
their superpartners. We will assume that these interactions 
together with the usual MSSM interactions are in equilibrium 
at high temperatures. The equilibrium number density of 
ultrarelativistic particles 
$\Delta n \equiv n_{{\rm{part}}} - n_{{\rm{antipart}}}$ 
is given by
\begin{equation}
\frac {\Delta n} {s} = \frac {15 g}{4 \pi^{2} g_{*}}
\left (\frac {\mu}{T}\right) 
\epsilon~,
\label{eq:chemical}
\end{equation}
where $g$ is the number of internal degrees of freedom of 
the particle under consideration, $\mu$ its chemical 
potential and $\epsilon=2$ or 1 for bosons or fermions. 
For each interaction in equilibrium, 
the algebraic sum of the chemical potentials of the particles 
involved is zero. Solving these constraints, we end up with 
only two independent chemical potentials, $\mu_{q} $ and 
$\mu_{\tilde{g}}$, and the $B$- and $L$- asymmetries 
are expressed \cite{ibanez} in terms of them:
$$
\frac {n_{B}}{s} = \frac {30}{4 \pi^{2}g_{*}T}
(6n_{g} \mu_{q} - (4n_{g} - 9) \mu_{\tilde{g}})~,
$$
\begin{equation}
\frac {n_{L}}{s} = - \frac {45}{4 \pi^{2} g_{*}T}
\left(\frac {n_{g}(14 n_{g} +9)}
{1+2 n_{g}} \mu_{q} + \Omega (n_{g}) 
\mu_{\tilde{g}}\right)~,
\label{eq:bla}
\end{equation}
where $\Omega(n_{g})$ is a known \cite{ibanez} function.
Now soft supersymmetry breaking couplings come in 
equilibrium at $T \stackrel{_{<}}{_{\sim }}10^7$ GeV 
since their rate $\Gamma_{S} \approx m^{2}_{3/2} /T 
\stackrel{_{>}}{_{\sim }} H \approx 30~T^{2}/M_{P}$. 
In particular, the nonvanishing gaugino mass implies 
$\mu _{\tilde{g}} =0$ and Eqs.(\ref{eq:bla}) give 
\cite{ibanez}
\begin{equation}
\frac {n_{B}}{s} = \frac {4(1+2n_{g})}{22n_{g}+13}
~\frac {n_{B-L}}{s}~.
\label{eq:bbminl}
\end{equation}
Equating $n_{B-L}/s$ with the primordial $n_{L}/s $, 
we have $n_{B}/s = (- 28/79) (n_{L}/s)$, for $n_{g}=3$.

\section {The `Left-Right' Model} 
\label{sec:lr}

\par
We will now study in detail a moderate extension of 
MSSM based on the left-right symmetric gauge group $G_{LR}$ 
which provides \cite{lss} a suitable framework for hybrid 
inflation. The inflaton is associated with the breaking of 
$SU(2)_{R}$ and consists of a gauge singlet and a pair of 
$SU(2)_{R}$ doublets. The $\mu$ problem is resolved 
\cite{dls} by introducing \cite{lss,dls} a trilinear 
superpotential coupling of the gauge singlet inflaton 
to the electroweak higgs doublets. In the presence of 
gravity-mediated supersymmetry breaking, this gauge singet 
acquires a vev and, thus, generates \cite{dls}, 
via its coupling to the higgses, the $\mu$ 
term.

\par     
The inflaton system, after the end of inflation, 
predominantly decays into higgs superfields 
and `reheats' the universe. Moreover, its subdominant 
decay into right handed neutrinos provides \cite{atmo} 
a mechanism for baryogenesis via leptogenesis. For 
$\nu_{\mu}$, $\nu_{\tau}$ masses from the small angle 
MSW resolution of the solar neutrino puzzle and the 
recent results of the SuperKamiokande experiment 
\cite{superk}, maximal $\nu_{\mu}-\nu_{\tau}$ mixing 
can be achieved \cite{atmo}. 

\subsection {The $\mu$ Problem} 
\label{subsec:mu}

\par
The breaking of $SU(2)_R\times U(1)_{B-L}$ is achieved 
by the renormalizable superpotential 
\begin{equation}
W = \kappa S(l^c\bar l^{c}- M^2)~, 
\label{W}
\end{equation}
where $S$ is a gauge singlet chiral superfield and 
$l^c$, $\bar l^{c}$ is a conjugate pair of $SU(2)_R$ 
doublet chiral superfields which acquire superheavy 
vevs of magnitude $M$. The parameters $\kappa$ and $M$ 
can be made positive by phase redefinitions.

\par
The $\mu$ problem can be resolved \cite{dls} by 
introducing the extra superpotential coupling
\begin{equation}
\delta W = \lambda S  h^2 =\lambda S \epsilon^{ij}
h_i^{(1)}h_j^{(2)}~,
\label{lambda} 
\end{equation}
where the chiral electroweak higgs superfield 
$h=(h^{(1)}, h^{(2)})$ belongs to a 
$(1,2,2)_{0}$ representation of $G_{LR}$ and 
$\lambda$ can again be made positive. The scalar 
potential which results from the  
terms in Eqs.(\ref{W}) and (\ref{lambda}) is 
(for canonical K\"ahler potential):
\begin{eqnarray}
V= | \kappa l^c\bar l^{c} + \lambda h^2 - 
\kappa M^2|^2 +(m_{3/2}^2 + \kappa^2 |\bar l^c|^2
+ \kappa^2 |l^c|^2 + \lambda^2 |h|^2)|S|^2 + 
m_{3/2}^2(|\bar l^c|^2 
\nonumber\\
+ |l^c|^2 + |h|^2) +  
\left (Am_{3/2}S( \kappa l^c\bar l^{c} + 
\lambda h^2 - \kappa M^2) + 2\kappa m_{3/2} M^{2}S+ 
{\rm h. c.} \right )~, 
\label{V}
\end{eqnarray}
where $m_{3/2}$ is the universal scalar mass (gravitino 
mass) and $A$ the universal coefficient of the trilinear
soft terms. For exact supersymmetry ($m_{3/2}
\rightarrow 0$), the vacua are \cite{dls} at
\begin{equation}
S = 0,~~~\kappa l^c\bar l^{c} + \lambda h^2 = 
\kappa M^2,~~~
l^c = e^{i\phi}\bar l^{c*}~~~h^{(1)}_i = 
e^{i\theta}\epsilon_{ij}h^{(2)j*},
\end{equation}
where the last two conditions arise from the requirement 
of D flatness. We see that there is a twofold degeneracy 
of the vacuum which is lifted by supersymmetry breaking. 
We get two degenerate (up to $m_{3/2}^{4}$) ground 
states ($\kappa\neq\lambda$): the desirable (`good') 
vacuum at $h = 0$ and $l^c\bar l^{c} = M^2$ and the 
undesirable (`bad') one at $h \neq 0$ and 
$l^c\bar l^{c} = 0$. They are separated by a potential 
barrier of order $M^{2}m_{3/2}^{2}~$.

\par
To leading order in supersymmetry breaking, the term of
the potential $V$ in Eq.(\ref{V}) proportional to $A$ 
vanishes, but a destabilizing tadpole term for $S$ 
remains:
\begin{equation} 
2\kappa m_{3/2}M^{2}S + {\rm h.c.}~.
\label{tadpole}
\end{equation} 
This term together with the mass term of $S$ 
(evaluated at the `good' vacuum) give 
$\langle S\rangle \approx - m_{3/2}/\kappa$ which, 
substituted in Eq.(\ref{lambda}), generates \cite{dls} 
a $\mu$ term with
\begin{equation}  
\mu =\lambda \langle S\rangle \approx - 
\frac{\lambda}{\kappa}m_{3/2}~.
\label{mu}
\end{equation}
Thus, coupling $S$ to the higgses can lead to the 
resolution of the $\mu$ problem.

\par
The model can be extended \cite{dls} to include matter 
fields too. The superpotential has the most general form 
respecting the $G_{LR}$ gauge symmetry and a global 
$U(1)$ R-symmetry. Baryon number is automatically 
implied by this R-symmetry to all orders in the 
superpotential, thereby guaranteeing the stability of 
proton.

\subsection{The Inflationary Trajectory}
\label{subsec:inf}

\par
The model has \cite{lss,dls} a built-in inflationary 
trajectory parametrized by $|S|$, $|S| > S_c=M$ for 
$\lambda>\kappa$ (see below). All other fields vanish 
on this trajectory. The $F_S$ term is constant providing 
a constant tree level vacuum energy density 
$\kappa^2 M^4$, which is responsible for 
inflation. One-loop radiative corrections (from the mass 
splitting in the supermultiplets $l^c$, $\bar l^{c}$ 
and $h$) generate a logarithmic slope \cite{dss} along 
the inflationary trajectory which drives the inflaton 
toward the minimum. For $|S| \leq S_c=M$, the $l^c$, 
$\bar l^{c}$ components become tachyonic and the system 
evolves towards the `good' supersymmetric minimum at 
$h=0$, $l^c=\bar l^c=M$ (for $\kappa > \lambda$, $h$ 
is destabilized first and the system would have evolved 
towards the `bad' minimum at $h \neq 0$, $l^c=
\bar l^c = 0$). For all values of the parameters 
considered here, inflation continues at least till $|S|$ 
approaches the instability at $|S|=S_c$ as one deduces 
from the slow roll conditions \cite{dss,lazarides}. The 
cosmic microwave quadrupole anisotropy can be calculated 
\cite {dss} by standard methods and turns out to be   
\begin{equation}
\left(\frac{\delta T}{T}\right)_{Q} \approx 
\frac{32 \pi^{5/2}}
{3\sqrt{5}}\left(\frac{M}{M_P}\right)^{3}
\kappa^{-1}x_{Q}^{-1}\Lambda (x_{Q})^{-1}~,
\label{anisotropy}
\end{equation}
\begin{eqnarray*}
\Lambda(x)=\left(\frac{\lambda}
{\kappa}\right)^{3}\left[\left(\frac{\lambda}
{\kappa}x^2-1\right)\ln \left(1-\frac{\kappa}
{\lambda}x^{-2}\right)
+\left(\frac{\lambda}{\kappa}x^2+1\right)
\ln \left(1+\frac{\kappa}
{\lambda}x^{-2}\right)\right] 
\end{eqnarray*}
\begin{equation}
+(x^2-1)\ln (1-x^{-2})+(x^2+1)\ln (1+x^{-2})~,
\label{temp}
\end{equation}
with $x=|S|/S_c$ and $S_Q$ being the value of $|S|$ 
when the present horizon scale crossed outside the 
inflationary horizon. (Notice that here we had to 
replace the contribution to the effective potential 
in Eq.(\ref{eq:veffexact}) from the $\phi$, 
$\bar{\phi}$ supermultiplets of the model in 
Sec.\ref{subsec:susy} by the contribution from the 
$l^c$, $\bar l^{c}$ and $h$ supermultiplets.) The 
number of e-foldings experienced by the universe between 
the time the quadrupole scale exited the horizon and the 
end of inflation is 
\begin{equation} 
N_Q \approx 32 \pi^3 \left(\frac{M}{M_P}\right)^2 
\kappa^{-2}\int_{1}^{x_{Q}^{2}}\frac{dx^2}{x^2}
\Lambda(x)^{-1}~.
\label{efoldings}
\end{equation}
The spectral index of density perturbations turns out 
to be very close to unity.

\subsection{`Reheating' and Leptogenesis}
\label{subsec:reheatlepto}

\par
After reaching the instability at $|S|=S_c$, the system 
continues \cite{bl} inflating for another e-folding or 
so reducing its energy density by a factor of about $2-3$~. 
It then rapidly settles into a regular oscillatory phase 
about the vacuum. Parametric resonance is safely ignored 
in this case \cite{bl}. The inflaton (oscillating system) 
consists of the two complex scalar fields $S$ and 
$\theta=(\delta \phi + \delta\bar{\phi})/\sqrt{2}$, 
where $\delta \phi = \phi - M$, $\delta \bar{\phi} = 
\bar{\phi} - M$, with mass $m_{infl}=\sqrt{2}\kappa M$ 
($\phi$, $\bar{\phi}$ are the neutral components of 
$l^c$, $\bar l^{c}$). 

\par
The scalar fields $S$ and $\theta$ predominantly decay 
into electroweak higgsinos and higgses respectively with 
a common decay width $\Gamma_{h}=(1/16\pi)\lambda^{2}
m_{infl}$, as one can easily deduce from the couplings 
in Eqs.(\ref{W}) and (\ref{lambda}). Note, however, 
that $\theta$ can also decay to right handed neutrinos 
$\nu^c$ through the nonrenormalizable superpotential 
term
\begin{equation} 
\frac{M_{\nu^c}}{2M^{2}}\bar{\phi} 
\bar{\phi} \nu^c \nu^c~,
\label{majorana}
\end{equation} 
allowed by the gauge and R- symmetries of the model 
\cite{lss,dls}. Here, $M_{\nu^c}$ denotes the Majorana 
mass of the relevant $\nu^c$. The scalar $\theta$ decays 
preferably into the heaviest $\nu^c$ with $M_{\nu^{c}} 
\leq m_{infl}/2$. The decay rate is given by
\begin{equation}
\Gamma_{\nu^c} \approx \frac{1}{16\pi}~\kappa^2 
m_{infl}~ \alpha^2 (1-\alpha^2)^{1/2}~,
\label{decayneu}
\end{equation}
where $0\leq \alpha=2M_{\nu^c}/m_{infl} \leq 1$.
The subsequent decay of these $\nu^{c}$ 's produces 
a primordial lepton number \cite{lepto} which is then 
partially converted to the observed BAU through 
electroweak sphaleron effects.
 
\par
The energy densities $\rho_{S}$, $\rho_{\theta}$, 
and $\rho_{r}$ of the oscillating fields $S$, $\theta$, 
and the `new' radiation produced by their decay to 
higgsinos, higgses and $\nu^c$ 's are controlled by the 
equations:
\begin{equation}
\dot{\rho}_{S}=-(3H+\Gamma_{h})\rho_{S}~,
~\rho_{\theta}(t)=\rho_{S}(t)
e^{-\Gamma_{\nu^c}(t-t_0)}~,
\label{infdensity}
\end{equation}
\begin{equation}
\dot{\rho}_{r}=-4H\rho_{r}+\Gamma_{h}\rho_{S}+
(\Gamma_{h}+\Gamma_{\nu^{c}})\rho_{\theta}~,
\label{raddensity}
\end{equation}
where
\begin{equation}
H=\frac{\sqrt{8\pi}}{\sqrt{3}M_P}~(\rho_{S}+
\rho_{\theta}+\rho_{r})^{1/2}
\label{hubble}
\end{equation}
is the Hubble parameter and overdots denote derivatives 
with respect to cosmic time $t$. The cosmic time at the 
onset of oscillations is taken $t_0\approx 0$. The 
initial values of the various energy densities are taken 
to be $\rho_{S}(t_0)=\rho_{\theta}(t_0)\approx 
\kappa^{2}M^{4}/6$, $\rho_{r}(t_{0})=0$. The `reheat' 
temperature $T_{r}$ is calculated from the equation
\begin{equation}
\rho_{S}+\rho_{\theta}=\rho_{r}=
\frac{\pi^2}{30}~g_{*}T_{r}^{4}~,
\label{reheat}
\end{equation}
where the effective number of massless degrees of 
freedom is $g_{*}$=228.75 for MSSM.

\par
The lepton number density $n_{L}$ produced by the 
$\nu^{c}$ 's satisfies the evolution equation:
\begin{equation}
\dot{n}_{L}=-3Hn_{L}+2\epsilon\Gamma_{\nu^c}
n_{\theta}~,
\label{lepton}
\end{equation}
where $\epsilon$ is the lepton number produced per 
decaying right handed neutrino and the factor of 2 in 
the second term of the rhs comes from the fact that we
get two $\nu^c$ 's for each decaying scalar $\theta$ 
particle. The `asymptotic' ($t\rightarrow 0$) lepton 
asymmetry turns out to be
\begin{equation}
\frac{n_{L}(t)}{s(t)}\sim 3
\left(\frac{15}{8}\right)^{1/4}\pi^{-1/2}
g_{*}^{-1/4}m_{infl}^{-1}~
\frac{\epsilon\Gamma_{\nu^c}}
{\Gamma_{h}+\Gamma_{\nu^c}}~\rho_{r}^{-3/4}
\rho_{S}e^{\Gamma_{h}t}~.
\label{leptonasymmetry}
\end{equation} 
 
\par
Assuming hierarchical light neutrino masses, we take 
$m_{\nu_{\mu}}\approx 2.6\times 10^{-3}~\rm{eV}$ 
which is the central value of the $\mu$-neutrino mass 
coming from the small angle MSW resolution of the solar 
neutrino problem \cite{smirnov}. The $\tau$-neutrino 
mass is taken $m_{\nu _{\tau}}\approx 7\times 10^{-2}
~\rm{eV}$, the central value from SuperKamiokande 
\cite{superk}. Recent analysis \cite{giunti} of 
the results of the CHOOZ experiment shows that 
the oscillations of solar and atmospheric neutrinos 
decouple. We thus concentrate on the two heaviest 
families ignoring the first one. Under these 
circumstances, the lepton number generated per 
decaying $\nu^c$ is \cite{lazarides,neu}
\begin{equation}
\epsilon=\frac{1}{8\pi}~g
\left(\frac{M_3}{M_2}\right)
~\frac{{\rm c}^{2}{\rm s}^{2}\ 
\sin 2\delta \ 
(m_{3}^{D}\,^{2}-m_{2}^{D}\,^{2})^{2}}
{|\langle h^{(1)}\rangle|^{2}~(m_{3}^{D}\,^{2}\ 
{\rm s}^{2}\ +
\ m_{2}^{D}\,^{2}{\rm \ c^{2}})}~,
\label{epsilon}
\end{equation}
where $g(r)=r\ln (1 + r^{-2})$~, 
$|\langle h^{(1)}\rangle|\approx 174~\rm{GeV}$, 
${\rm c}=\cos \theta ,\ {\rm s}=\sin \theta $, 
and $\theta$ ($0\leq \theta \leq \pi /2$) and 
$\delta$ ($-\pi/2\leq \delta <\pi/2 $) are the 
rotation angle and phase which diagonalize the 
Majorana mass matrix of $\nu^{c}$ 's with eigenvalues 
$M_2$, $M_3$ ($\geq 0$). The `Dirac' mass matrix 
of the neutrinos is considered diagonal with 
eigenvalues $m_{2}^{D}$, $m_{3}^{D}$ ($\geq 0$). 

\par
For the range of parameters considered here, the 
scalar $\theta$ decays into the second heaviest right 
handed neutrino with mass $M_{2}$ ($<M_{3}$) and, 
thus, $M_{\nu^{c}}$ in Eqs.(\ref{majorana}) and 
(\ref{decayneu}) should be 
identified with $M_{2}$. Moreover, $M_{3}$ turns out 
to be bigger than $m_{infl}/2$ as it should. We will 
denote the two positive eigenvalues of the light neutrino 
mass matrix by $m_{2}$ (=$m_{\nu _{\mu }}$), $m_{3}$ 
(=$m_{\nu _{\tau }}$) with $m_{2}\leq m_{3}$. All the 
quantities here (masses, rotation angles and phases) are 
`asymptotic' (defined at the grand unification scale 
$M_{GUT}$). 

\par
The determinant and the trace invariance of 
the light neutrino mass matrix imply\cite{neu} 
two constraints on the (asymptotic) parameters which 
take the form: 
\begin{equation}
m_{2}m_{3}\ =\ \frac{\left( m_{2}^{D}m_{3}^{D}
\right) ^{2}}{M_{2}\ M_{3}}~,
\label{determinant}
\end{equation}
\begin{eqnarray*}
m_{2}\,^{2}+m_{3}\,^{2}\ =\frac{\left( m_{2}^{D}\,\,
^{2}{\rm c}^{2}+m_{3}^{D}\,^{2}{\rm s}^{2}\right) 
^{2}}{M_{2}\,^{2}}+
\end{eqnarray*}
\begin{equation}
\ \frac{\left( m_{3}^{D}\,^{2}{\rm c}^{2}+m_{2}^{D}\,
^{2}{\rm s}^{2}\right)^{2}}{M_{3}\,^{2}}+\ 
\frac{2(m_{3}^{D}\,^{2}-m_{2}^{D}\,^{2})^{2}
{\rm c}^{2}{\rm s}^{2}\,{\cos 2\delta }}
{M_{2}\,M_{3}}~\cdot
\label{trace} 
\end{equation}

\par
The $\mu-\tau$ mixing angle $\theta _{23}$ 
(=$\theta _{\mu\tau}$) lies \cite{neu} in the range
\begin{equation}
|\,\varphi -\theta ^{D}|\leq \theta _{23}\leq
\varphi +\theta ^{D},\ {\rm {for}\ \varphi +
\theta }^{D}\leq \ \pi /2~,~~~~~
\label{mixing}
\end{equation}
where $\varphi$ ($0\leq \varphi \leq \pi /2$) is the 
rotation angle which diagonalizes the light neutrino mass 
matrix, and $\theta ^{D}$ ($0\leq \theta ^{D} \leq 
\pi /2$) is the `Dirac' (unphysical) mixing angle in 
the $2-3$ leptonic sector defined in the absence of the 
Majorana masses of the $\nu^{c}$ 's.

\par
Assuming approximate $SU(4)_{c}$ symmetry, we get the 
asymptotic (at $M_{GUT}$) relations: 
\begin{equation}
m_{2}^{D}\approx m_{c}\ ,
\ m_{3}^{D}\approx \ m_{t}\ ,
\ \sin\theta ^{D}\approx |V_{cb}|~.
\label{asympt}
\end{equation}
Renormalization effects, for MSSM spectrum and  
$\tan \beta \approx m_{t}/m_{b}$, are incorporated 
\cite{neu} by substituting in the above 
formulas the values: $m_{2}^{D}\approx 0.23~{\rm GeV}$, 
$\ m_{3}^{D}\approx 116$ GeV and $\sin \theta ^{D}
\approx 0.03$. Also, $\tan^{2} 2 \theta _{23}$ 
increases by about 40\% from $M_{GUT}$ to $M_{Z}$.

\par
We take a specific MSSM framework \cite{als} 
where the three Yukawa couplings of the third generation 
unify `asymptotically' and, thus, 
$\tan \beta \approx m_{t}/m_{b}$. We choose the 
universal scalar mass (gravitino mass) $m_{3/2} 
\approx 290~{\rm{GeV}}$ and the universal gaugino mass 
$M_{1/2} \approx 470~{\rm{GeV}}$. These values 
correspond \cite{asw} to $m_{t}(m_{t})\approx 166~
{\rm{GeV}}$ and $m_{A}$ (the tree level CP-odd scalar 
higgs mass) =$M_{Z}$. The ratio $\lambda/\kappa$ is 
evaluated \cite{carena} from 
\begin{equation}
\frac{\lambda}{\kappa}=\frac{|\mu|}{m_{3/2}}\approx 
\frac{M_{1/2}}{m_{3/2}}\left(1-\frac{Y_{t}}
{Y_{f}}\right)^{-3/7}\approx 3.95~,
\label{renorm}
\end{equation} 
where $Y_{t}=h_{t}^2\approx 0.91$ is the square of the 
top-quark Yukawa coupling and $Y_{f}\approx 1.04$ is the 
weak scale value of $Y_{t}$ corresponding to `infinite' 
value at $M_{GUT}$.

\par
\begin{figure}
\epsfig{figure=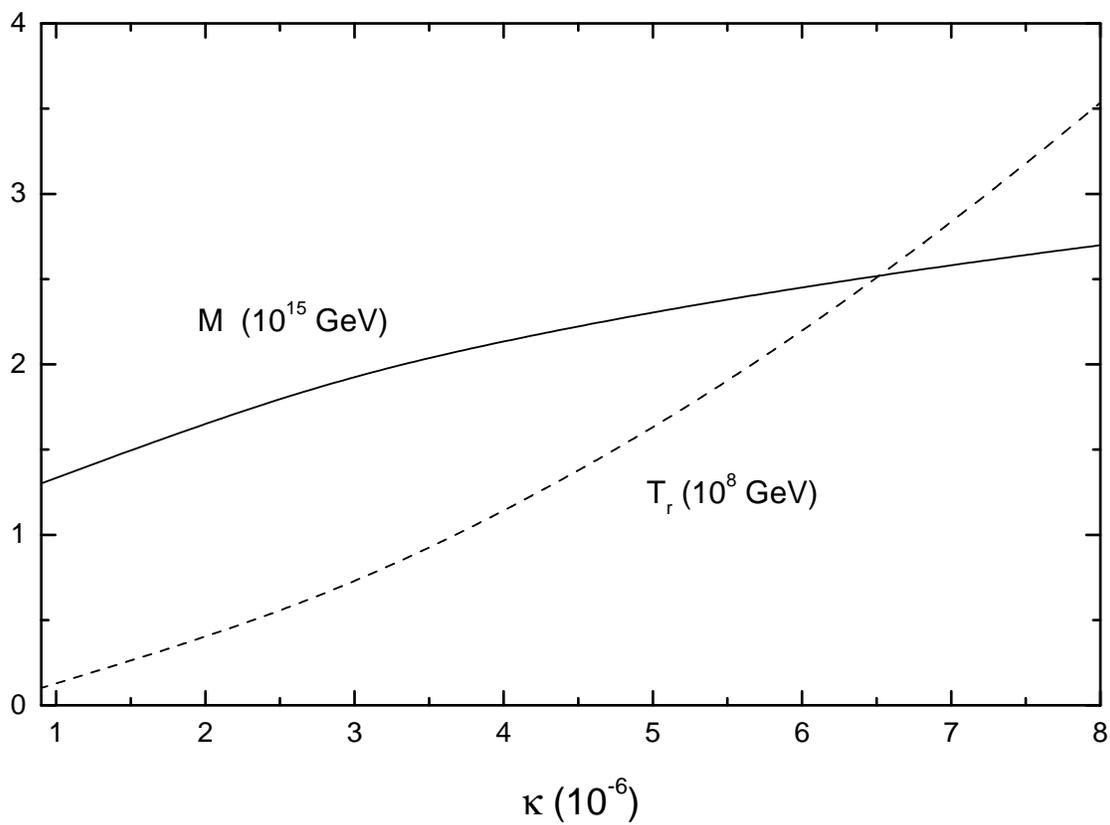,height=5.8in,angle=-90}
\medskip
\caption{The mass scale $M$ (solid line) and the reheat 
temperature $T_{r}$ (dashed line) as functions of 
$\kappa$.
\label{reheating}}
\end{figure}

Eqs.(\ref{anisotropy})-(\ref{efoldings}) can now be 
solved, for $(\delta T/T)_{Q} \approx 6.6\times 10^{-6}$ 
from COBE, $N_Q \approx 50$ and any value of $x_{Q}>1$. 
Eliminating $x_{Q}$, we obtain $M$ as a function of 
$\kappa$ depicted in Fig.\ref{reheating}. The evolution 
Eqs.(\ref{infdensity})-(\ref{hubble}) are solved 
for each value of $\kappa$. The parameter $\alpha^{2}$ 
in Eq.(\ref{decayneu}) is taken equal to 2/3. This 
choice maximizes the decay width of the inflaton 
to $\nu^{c}$ 's and, thus, the subsequently produced 
lepton asymmetry. The `reheat' temperature is then 
calculated from Eq.(\ref{reheat}) for each value of 
$\kappa$. The result is again depicted in 
Fig.\ref{reheating}.

\par
\begin{figure}
\epsfig{figure=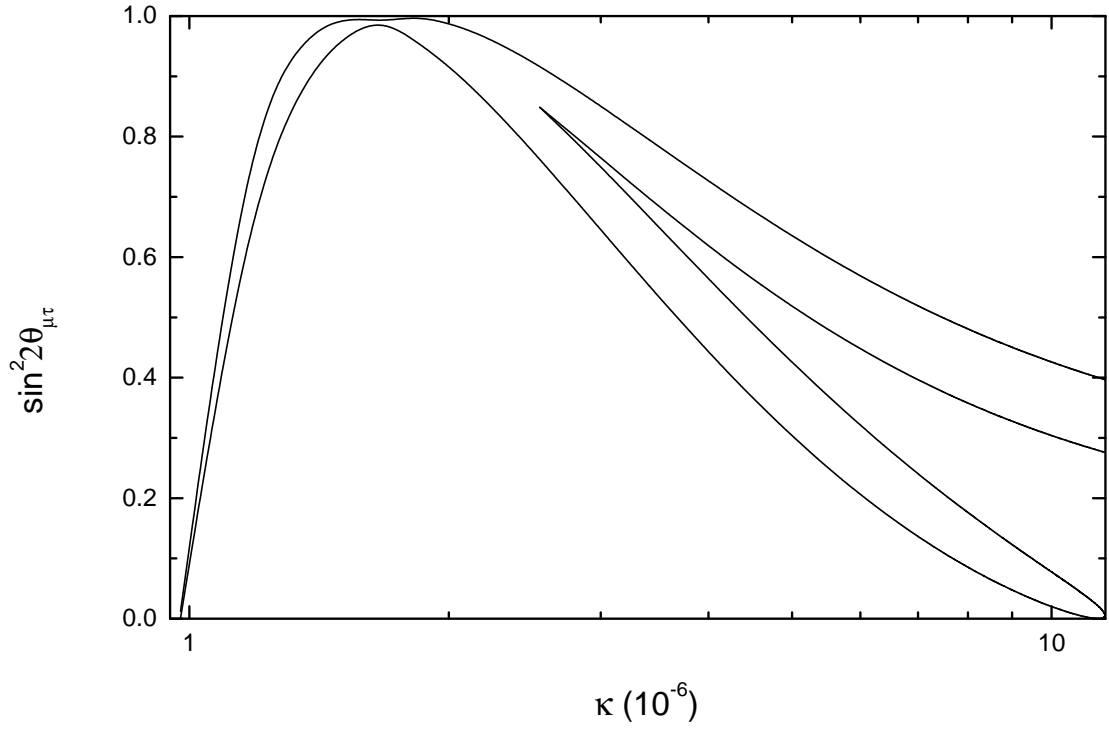,height=5.8in,angle=-90}
\medskip
\caption{The allowed region (bounded by the solid lines)
in the $\kappa-\sin^{2} 2 \theta _{\mu\tau}$ plane 
for $m_{\nu_{\mu}}\approx 2.6\times 10^{-3}~\rm{eV}$ 
and $m_{\nu_{\tau}}\approx 7 \times 10^{-2}~\rm{eV}$.
\label{angle}}
\end{figure}

 
The mass of the second heaviest $\nu^c$, into which the 
scalar $\theta$ decays partially, is given by $M_{2}=
M_{\nu^{c}}=\alpha m_{infl}/2$ and $M_{3}$ is found 
from the `determinant' condition in Eq.(\ref{determinant}). 
The `trace' condition in Eq.(\ref{trace}) is then solved 
for $\delta(\theta )$ which is subsequently substituted 
in Eq.(\ref{epsilon}) for $\epsilon$. The leptonic 
asymmetry as a function of the angle $\theta$ can be found 
from Eq.(\ref{leptonasymmetry}). For each value of 
$\kappa$, there are two values of $\theta$ satisfying the 
low deuterium abundance constraint $\Omega _{B}h^{2}
\approx 0.025$. (These values of $\theta$ turn out to be 
quite insensitive to the exact value of $n_{B}/s$.) The 
corresponding $\varphi$ 's are then found and the allowed 
region of the mixing angle $\theta _{\mu\tau}$ 
in Eq.(\ref{mixing}) is determined for each $\kappa$. 
Taking into account renomalization effects and 
superimposing all the permitted regions, we obtain the 
allowed range of $\sin^{2} 2 \theta _{\mu\tau}$ 
as a function of $\kappa$, shown in Fig.\ref{angle}. 
We observe that $\sin^{2} 2 \theta _{\mu\tau}
\stackrel{_{>}}{_{\sim }} 0.8$ (from SuperKamiokande 
\cite{superk}) corresponds to $1.2\times 10^{-6}
\stackrel{_{<}}{_{\sim }}\kappa
\stackrel{_{<}}{_{\sim }}3.4\times 10^{-6}$ which 
is rather small. (Fortunately, supersymmetry protects it 
from radiative corrections.) 

\par
The corresponding values of $M$ and $T_{r}$ can be read 
from Fig.\ref{reheating}. One finds that $1.4\times 10^{15}
~{\rm{GeV}}\stackrel{_{<}}{_{\sim }} M
\stackrel{_{<}}{_{\sim }}2\times 10^{15}~{\rm{GeV}}$ 
and $1.8\times 10^{7}~{\rm{GeV}}\stackrel{_{<}}{_{\sim }} 
T_{r}\stackrel{_{<}}{_{\sim }}8.7\times 10^{7}
~{\rm{GeV}}$. We observe that $M$ turns out to be somewhat 
smaller than the MSSM unification scale $M_{GUT}$. (It is 
anticipated that $G_{LR}$ is embedded in a grand unified 
theory.) The `reheat' temperature, however, satisfies the 
gravitino constraint 
($T_{r}\stackrel{_{<}}{_{\sim}} 10^{9}~{\rm{GeV}}$).
Note that, for the values of the parameters
chosen here, the lightest supersymmetric particle (LSP)
is \cite{asw} an almost pure bino with mass $m_{LSP}
\approx 0.43M_{1/2}\approx 200~{\rm{GeV}}$
and can, in principle, provide the cold dark matter of the 
universe. On the contrary, there is no hot dark matter 
candidate, in the simplest scheme.

\par
In conclusion, we have shown that, in a supersymmetric 
model based on a left-right symmetric gauge group and 
leading `naturally' to hybrid inflation, the $\mu$ 
problem can be easily solved . The observed BAU is 
produced via a primordial leptogenesis. For masses of 
$\nu_{\mu}$, $\nu_{\tau}$ from the small angle MSW 
resolution of the solar neutrino puzzle and SuperKamiokande, 
maximal $\nu_{\mu}-\nu_{\tau}$ mixing can be achieved. 
The required values of the coupling constant $\kappa$ are, 
however, quite small ($\sim 10^{-6}$).
 
\vspace{0.5cm}
This work is supported by E.U. under TMR contract 
No. ERBFMRX--CT96--0090.

\def\ijmp#1#2#3{{ Int. Jour. Mod. Phys. }
{\bf #1~}(19#2)~#3}
\def\pl#1#2#3{{ Phys. Lett. }{\bf B#1~}(19#2)~#3}
\def\zp#1#2#3{{ Z. Phys. }{\bf C#1~}(19#2)~#3}
\def\prl#1#2#3{{ Phys. Rev. Lett. }{\bf #1~}(19#2)~#3}
\def\rmp#1#2#3{{ Rev. Mod. Phys. }{\bf #1~}(19#2)~#3}
\def\prep#1#2#3{{ Phys. Rep. }{\bf #1~}(19#2)~#3}
\def\pr#1#2#3{{ Phys. Rev. }{\bf D#1~}(19#2)~#3}
\def\np#1#2#3{{ Nucl. Phys. }{\bf B#1~}(19#2)~#3}
\def\mpl#1#2#3{{ Mod. Phys. Lett. }{\bf #1~}(19#2)~#3}
\def\arnps#1#2#3{{ Annu. Rev. Nucl. Part. Sci. }{\bf
#1~}(19#2)~#3}
\def\sjnp#1#2#3{{ Sov. J. Nucl. Phys. }{\bf #1~}
(19#2)~#3}
\def\jetp#1#2#3{{ JETP Lett. }{\bf #1~}(19#2)~#3}
\def\app#1#2#3{{ Acta Phys. Polon. }{\bf #1~}(19#2)~#3}
\def\rnc#1#2#3{{ Riv. Nuovo Cim. }{\bf #1~}(19#2)~#3}
\def\ap#1#2#3{{ Ann. Phys. }{\bf #1~}(19#2)~#3}
\def\ptp#1#2#3{{ Prog. Theor. Phys. }{\bf #1~}
(19#2)~#3}
\def\plb#1#2#3{{ Phys. Lett. }{\bf#1B~}(19#2)~#3}
\def\ibid#1#2#3{{ ibid. }{\bf#1~}(19#2)~#3}
\def\apjl#1#2#3{{ Astrophys. J. Lett. }{\bf#1~}(19#2)~#3}

\newpage

\pagestyle{empty}

\begin{references}

\bibitem{hybrid} A. D. Linde, \pl{259}{91}{38}; 
\pr{49}{94}{748}.

\bibitem{cobe} G. F. Smoot {\it et al.}, \apjl{396}{92}
{L1}; C. L. Bennett {\it et al.},  \apjl{464}{96}{1}.

\bibitem{lyth} E. J. Copeland, A. R. Liddle, D. H. Lyth, 
E. D. Stewart and D. Wands, \pr{49}{94}{6410}.

\bibitem{dss} G. Dvali, Q. Shafi and R. Schaefer, 
\prl{73}{94}{1886}.

\bibitem{lss} G. Lazarides, R. Schaefer and Q. Shafi, 
\pr{56}{97}{1324}.

\bibitem{dls} G. Dvali, G. Lazarides and Q. Shafi, 
\pl{424}{98}{259}.

\bibitem{atmo} G. Lazarides and N. D. Vlachos, \pl{441}
{98}{46}.

\bibitem{superk} T. Kajiata, talk given at the XVIIIth 
International Conference on Neutrino Physics and 
Astrophysics (Neutrino'98), Takayama, Japan, 4-9 June, 1998.

\bibitem{new} A. D. Linde, \plb{108}{82}{389}; 
A. Albrecht and P. Steinhardt, \prl{48}{82}{1220}.

\bibitem{chaotic} A. D. Linde, \jetp{38}{83}{149}; 
\plb{129}{83}{177}.

\bibitem{lp} G. Lazarides and C. Panagiotakopoulos, 
\pr{52}{95}{R559}.

\bibitem{lazarides} G. Lazarides, hep-ph/9802415 (lectures 
given at the 6th BCSPIN Summer School).

\bibitem{sugra} G. Lazarides and N. Tetradis, \pr{58}
{98}{123502}; E. D. Stewart, \pr{51}{95}{6847}.

\bibitem{baryonparity}  Q. Shafi and X. M. Wang, 
unpublished(1990); L. Ibanez and G. G. Ross, 
\np{368}{92}{3}; G. Lazarides, C. Panagiotakopoulos 
and Q. Shafi, \pl{315}{93}{325}, (E)\ibid{317}{93}{661}.

\bibitem{trinifiedmonopoles} G. Lazarides, 
C. Panagiotakopoulos 
and Q. Shafi, \prl{58}{87}{1707}.

\bibitem{gravitino} M. Yu. Khlopov and A. D. Linde, 
\plb{138}{84}{265}; J. Ellis, J. E. Kim and 
D. Nanopoulos, \plb{145}{84}{181}; M. Kawasaki and 
T. Moroi, \ptp{93}{95}{879}. 

\bibitem{lepto} M. Fukugita and T. Yanagida, 
\pl{174}{86}{45}; 
W. Buchm\"uller and M. Pl\"umacher, \pl{389}{96}{73}. 
In the context of inflation see G. Lazarides and 
Q. Shafi, \pl{258}{91}{305}. 

\bibitem{sphaleron} S. Dimopoulos and L. Susskind, 
\pr{18}{78}{4500}; V. Kuzmin, V. Rubakov and 
M. Shaposhnikov, \pl{155}{85}{36}; P. Arnold and 
L. McLerran, \pr{36}{87}{581}.

\bibitem{turner} J. A. Harvey and M. S. Turner, 
\pr{42}{90}{3344}.

\bibitem{ibanez} L. E. Ib\'a\~nez and F. Quevedo, 
\pl{283}{92}{261}.

\bibitem{thooft} G.'t Hooft, \prl{37}{76}{8}; 
\pr{14}{76}{3432}.

\bibitem{vacuum} C. Callan, R. Dashen and D. Gross, 
\plb{63}{76}{334}; R. Jackiw and C. Rebbi, \prl{37}
{76}{172}.

\bibitem{instanton} A. A. Belavin, A. Polyakov, 
A. Schwartz and Y. Tyupkin, \plb{59}{75}{85}.

\bibitem{restricted} I. Affleck, \np{191}{81}{429}.

\bibitem{sphaleronsol} N. S. Manton, \pr{28}{83}{2019};
F. R. Klinkhamer and N. S. Manton, \pr{30}{84}{2212}.

\bibitem{bl} J. Garc\'ia-Bellido and A. Linde, 
\pr{57}{98}{6075}.

\bibitem{smirnov} A. Smirnov, hep-ph/9611465 and 
references therein.

\bibitem{giunti} C. Giunti, hep-ph/9802201.

\bibitem{neu} G. Lazarides, Q. Shafi and N. D. Vlachos, 
\pl{427}{98}{53}.

\bibitem{als} B. Ananthanarayan, G. Lazarides and Q. Shafi,
\pr{44}{91}{1613}.

\bibitem{asw} B. Ananthanarayan, Q. Shafi and X. M. Wang,
\pr{50}{94}{5980}.

\bibitem{carena} M. Carena, M. Olechowski, 
S. Pokorski and C.Wagner, \np{426}{94}{269}.
 
\end{references}
\end{document}